# Quantum Pressure Focusing in Solids:

## A Reconstruction from Experimental Electron Density

By Vladimir G. Tsirelson [a, b], Adam I. Stash[a, c] and Ilya V. Tokatly [d,e,f]

[a] *Quantum Chemistry Department, Mendeleev University of Chemical Technology, Moscow, 125047, Russia*
[b] *Chemistry Department, South Ural State University, Lenin Prospect, 76, Chelyabinsk, 454080, Russia*
[c] *Laboratory of oxide materials, Karpov Research Institute of Physical Chemistry, Moscow, 103064, Russia*
[d] *Departamento de Fisica de Materals, Universidad del Pais Vasco UPV/EHU,20018 San Sebastian, Spain*
[e] *IKERBASQUE, Basque Foundation for Science, E-20018, Bilbao, Spain*
[f] *Donostia International Physics Center (DIPC), Manuel de Lardizabal 5, E-20018 San Sebastián, Spain*

Corresponding author: Vladimir G. Tsirelson. E-mail vtsirelson@yandex.ru

## Abstract

We present an approach for reconstructing the distribution of electronic internal quantum pressure in the electronic continuum of solids from the experimental electron density. Using the formalism of the density functional theory, we obtain the spatial inner-crystal map of the quantum pressure. The results are visualized via the indicator of quantum pressure focusing (IQPF) which reveals the regions where the pressure is concentrated or depleted due to the quantum effects. IQPF contains all quantum electron shell structure-forming contributions resulting from kinetic, exchange and correlation effects, and presents a clear picture of the chemical bond features in crystals with different type of bonding mechanisms.

Electrons in atoms, molecules or solids reach the equilibrium state when the external force exerted from the nuclei on the electrons is balanced at each point of position space by the internal stress force $f_{int}(\mathbf{r}) = \nabla\boldsymbol{\sigma}_{int}(\mathbf{r})$ of the electronic subsystem, where $\boldsymbol{\sigma}_{int}(\mathbf{r})$ is the electron stress tensor (Nielsen and Martin, 1983, 1985; Folland, 1986;. Godfrey, 1988; Uspenskii, Ziesche & Gräfenstein, 1989; Gräfenstein &. Ziesche, 1996; Filippetti & Fiorentini, 2000; Rogers & Rappe, 2001, 2002; Tachibana, 2004, 2005, 2010, 2017; Maranganti, Sharma & Wheeler, 2007; Szarek, Sueda & Tachibana, 2008; Maranganti & Sharma. 2010; Anderson, Ayers, & Hernandez, 2010; Cohen, 2011; Dem'yanov & Polestshuk, 2012; Martín Pendás & Hernández-Trujillo, 2012; Finzel & Kohout, 2013; Maza, Jenkins, Kirk, et al, 2013; Finzel, 2014) . As the nuclei are kept together by electronic continuum, one may expect that the space distribution of the electron stress contains information about the chemical bonding. It looks indeed quite tempting to interpret the electronic shells of bonded atoms and resulting from them chemical bond patterns as compressed or stretched regions spontaneously formed in the electron continuum in the field of nuclei (Tao, Vignale, & Tokatly, 2008, 2015; Tsirelson, Stash & Tokatly, 2016).

In classical continuum mechanics, the stress tensor is usually defined via the work done against the internal forces (the change of the internal energy of a system) when the continuum is subject to an external deformation $\delta A = \int \sigma_{int, ij}(\mathbf{r})\varepsilon_{ij}(\mathbf{r})dV$ , where $\varepsilon_{ij}(\mathbf{r})$ is the strain tensor (Landau & Lifshitz, 1986). This definition can be easily extended to a generic quantum system by noticing a relation of the strain tensor to the change $\delta g_{ij}(\mathbf{r})$ of the metric tensor induced by the deformation, $\varepsilon_{ij}(\mathbf{r})=(1/2)\,\delta g_{ij}(\mathbf{r})$ (Fung, 1965). This relation allows to define the stress tensor by placing the system in a space with weakly distorted metrics $g_{ij}(\mathbf{r}) = \delta_{ij} + \delta g_{ij}(\mathbf{r})$ and computing the corresponding local change of the internal energy, i.e. calculating the variational derivative of the internal energy with respect to the metric tensor (Rogers & Rappe, 2002). In the above formulation, the stress tensor $\boldsymbol{\sigma}(\mathbf{r})$ appears as a natural local probe of energetics in a response to the deformation of an infinitesimal element of the electron fluid at a given point of space.

Unfortunately the geometric definition still suffers for the usual one-parameter ambiguity of the kinetic contribution to the stress (Filippetti & Fiorentini, 2000; Rogers & Rappe, 2002; Maranganti, Sharma, & Wheeler, 2007) , which can be traced back to the old and still unresolved problem of quantization in curved spaces and/or constrained surfaces (see e.g. Kaplan, Maitra,& Heller, 1997; Schuster & Jaffe,2003; Fiziev & Kleinert, 1996, and references therein). In the geometric formulation of the stress, the ambiguity shows up as the possibility to add to the Hamiltonian an effective one-particle potential proportional to the scalar curvature (Rogers & Rappe, 2002). This potential vanishes in the physical flat space, but it does contribute to the stress tensor. Importantly, the corresponding correction to the force vanishes identically and does not influence the physical uniqueness of the stress force.

In fact, the ambiguity of the stress tensor is similar to a gauge freedom (Landau & Lifshitz, 2007). In the situation when there is no a clear mathematical reason favoring a particular gauge, the choice can be made based on convenience/simplicity and a consistency with a physical and chemical thinking. Apparently the simplest choice is to set the coefficient in front of the scalar curvature to zero. This corresponds to a "minimal coupling" of the electrons to the deformation via the replacement of the Laplace operator in the kinetic energy expression by its deformed, covariant counterpart, the Laplace-Beltrami operator (Jost, 2002),. The stress tensor defined via the "minimal coupling" procedure naturally appears in the context of quantum dynamics in local non-inertial frames, such as a co-moving Lagrangian frame (Tokatly, 2005, 2007), or local frame that follows the nuclei displaced from their ideal positions in the geometric approach to flexoelectricity (Stengel & Vanderbilt, 2018). The latter example is especially relevant in the present context as it connects the local electronic stress to external deformations generated, for example in the high-pressure experiments (Freddi, Olmi, & Cristofolini, 2015).

Remarkably, the "minimal" geometric definition of the stress tensor used by Tao, Vignale, & Tokatly (2008) naturally lead to the idea of quantum stress focusing and the visualization of atomic electron shells and chemical bonds in terms of stressed regions in the electron continuum. In the present paper we also adopt the same "minimal" form of the electron stress tensor and its physical interpretation as a local change of internal energy in a response to deformations of infinitesimal electronic media elements.

Physically the electron stress originates from the kinetic and the electron-electron interaction effects. Note, that the mean-field Hartree contribution to the interaction part of the stress tensor does not carry significant bonding information: it is even more structureless then the electron density (Tsirelson, Avilov, Lepeshov, et al. 2001). Therefore, in the following we exclude the featureless Hartree contribution from the stress tensor and consider it as an external electrostatic potential together with the field from the nuclei.

The rest of the stress tensor, i.e. the kinetic and the exchange-correlation contributions of quantum nature, can be conveniently separated into the isotropic (pressure) part, $p(\mathbf{r})$, and the traceless (shear) part, $\pi_{ij}(\mathbf{r})$:

$$\sigma_{ij}(\mathbf{r}) = -\delta_{ij}p(\mathbf{r}) + \pi_{ij}(\mathbf{r}). \qquad (1)$$

The quantum pressure is defined as a negative of the trace of the stress tensor, $p(\mathbf{r}) = -(1/3)\,\mathrm{Tr}\boldsymbol{\sigma}(\mathbf{r})$. Physically the pressure $p(\mathbf{r})$ characterizes the variation of the internal energy under a local deformation that changes the volume of a small element of the electron continuum without changing its shape. In particular in the regions with a positive pressure, a local compression leads to the increase of the energy. These regions can thus be regarded as "compressed" (resisting the further compression). By the same token, the regions of a negative pressure are "stretched" (favoring the external local compression as it leads to the energy decrease). The shear part $\pi_{ij}(\mathbf{r})$ describes the anisotropy of the stress distribution. It portrays the response to a local reshaping of the electron continuum element at $\mathbf{r}$, while preserving its volume.

It is clear that the internal quantum pressure $p(\mathbf{r})$ carries only a part of the information about the stress of the electronic continuum. Nonetheless, it has been found that the position distribution of the pressure recovers the main chemical bond features in molecules and solids related to the spatial structure of electronic shells (Tao, Vignale, & Tokatly, 2008; Tsirelson, Stash & Tokatly. 2016). In this work we demonstrate that these features in solids may be obtained from the experimental electron density measured by the precise high-resolution X-ray diffraction (Tsirelson, & Ozerov, 1996; Tsirelson, 2007). Specifically we reconstruct a function that is obtained by properly rescaling the quantum pressure (Tao, Vignale & Tokatly, 2008; Tao, Liu, Zheng & Rappe, 2015) in a way that is somewhat similar to constructing the well-known electron localization function (ELF) by rescaling the local kinetic energy (Becke & Edgecombe, 1990; Savin, Becke, Flad, et al, 1991; Silvi & Savin, 1994). Our function $L(\mathbf{r})$ (see below) conveniently maps the quantum pressure and reveals the atomic shells and chemical bond features as the regions where the pressure is concentrated or depleted due to the quantum effects. In the following we call this function $L(\mathbf{r})$ as Indicator of Quantum Pressure Focusing (IQPF). We stress that IQPF contains all quantum electron shell structure-forming contributions resulting from kinetic, exchange and correlation effects, whereas the common ELF reflects only kinetic component of bonding.

**Method**

In the framework of density functional theory (DFT) (Dreizler & Gross, 1990), the quantum electron stress is partitioned into kinetic Kohn-Sham, $\sigma_{ij}^{S}(\mathbf{r})$, and exchange-correlation, $\sigma_{ij}^{xc}(\mathbf{r})$, parts:

$$\sigma_{ij}(\mathbf{r}) = \sigma_{ij}^{S}(\mathbf{r}) + \sigma_{ij}^{xc}(\mathbf{r}) . \qquad (2)$$

As we discussed in the Introduction, to define kinetic part of the quantum stress tensor we follow Tao, Vignale & Tokatly (2008) and adopt the minimal geometric coupling of electrons to the deformations, which originates from the description of quantum dynamics in local non-inertial frames (Tokatly, 2005, 2008). This definition allows for a clear physical interpretation of the local stress in terms of the internal energy response to a local deformation of small elements of the electron liquid. Explicitly the kinetic stress tensor reads as

$$\sigma_{ij}^{S}(\mathbf{r}) = -\left\{ \frac{1}{2} \sum_{l\sigma} \left[ \frac{\partial \psi_{l\sigma}^{*}(\mathbf{r})}{\partial r_i} \frac{\partial \psi_{l\sigma}(\mathbf{r})}{\partial r_j} + \frac{\partial \psi_{l\sigma}^{*}(\mathbf{r})}{\partial r_j} \frac{\partial \psi_{l\sigma}(\mathbf{r})}{\partial r_i} \right] - \frac{1}{4}\delta_{ij}\nabla^2\rho(\mathbf{r}) \right\}, \quad i,j = x,y,z \quad (3)$$

where $\psi_{l\sigma}(\mathbf{r})$ are Kohn-Sham spin-orbitals and $\rho(\mathbf{r})$ is electron density; atomic units are used throughout the paper. The exchange-correlation part of the quantum stress tensor, $\sigma_{ij}^{xc}(\mathbf{r})$, in the generalized gradient approximations of DFT has the form (Tao, Vignale & Tokatly, 2008)

$$\sigma_{ij}^{xc}(\mathbf{r}) = -\delta_{ij}[\rho(\mathbf{r})v_{xc}(\mathbf{r}) - e_{xc}(\mathbf{r})] - \frac{\dfrac{\partial\rho(\mathbf{r})}{\partial r_i}\dfrac{\partial\rho(\mathbf{r})}{\partial r_j}}{2k_F(\mathbf{r})\rho(\mathbf{r})\left|\nabla\rho(\mathbf{r})\right|}\frac{\partial e_{xc}(\mathbf{r})}{\partial s}\ . \qquad (4)$$

Here $v_{xc}(\mathbf{r})$ is electron exchange-correlation potential, $s(\mathbf{r}) = \left|\nabla\rho(\mathbf{r})\right| / 2k_F(\mathbf{r})\rho(\mathbf{r})$ is reduced density gradient, $k_F(\mathbf{r}) = (3\pi^2\rho(\mathbf{r}))^{1/3}$. The electron exchange-correlation energy density, $e_{xc}(\mathbf{r})$, is defined according to $E_{xc}[\rho] = \int e_{xc}(\mathbf{r})d\mathbf{r}$ (Dreizler & Gross , 1990)

By the definition of pressure, $p(\mathbf{r}) = -(1/3)\,\mathrm{Tr}\boldsymbol{\sigma}(\mathbf{r})$. Therefore the kinetic and exchange-correlation contributions to the quantum electron pressure is obtained by taking the trace of the corresponding parts of the stress tensor,

$$p^S(\mathbf{r}) = \frac{1}{3}\sum_{l\sigma}\left|\nabla\psi_{l\sigma}(\mathbf{r})\right|^2 - \frac{1}{4}\nabla^2\rho(\mathbf{r}), \quad i, j = x, y, z\ . \qquad (5)$$

$$p^{xc}(\mathbf{r}) = \rho(\mathbf{r})v_{xc}(\mathbf{r}) - e_{xc}(\mathbf{r}) + \frac{s}{3}\frac{\partial e_{xc}(\mathbf{r})}{\partial s}\ . \qquad (6)$$

Typically, the electron correlation contribution to the local stress is expected to be small. Therefore here we approximate the exchange-correlation part of the pressure by the exchange-only LDA expression

$$p_x(\mathbf{r}) = -\frac{1}{4}(\frac{3}{\pi})^{1/3}\rho^{4/3}(\mathbf{r})\ . \qquad (7)$$

Then the total local quantum pressure takes the following form:

$$p(\mathbf{r}) = p^S(\mathbf{r}) - \frac{1}{4}(\frac{3}{\pi})^{1/3}\rho^{4/3}(\mathbf{r})\ . \qquad (8)$$

In this work we aim at analyzing the internal pressure of inhomogeneous electron gas in a crystal using experimental electron density measured by the precise high-resolution X-ray diffraction (Tsirelson, & Ozerov, 1996). To this end we rewrite expression (5) in the form

$$p^S(\mathbf{r}) = \frac{2}{3} t_s(\mathbf{r}) - \frac{1}{4} \nabla^2 \rho(\mathbf{r}) \qquad (9)$$

where $t_s(\mathbf{r})$ is kinetic energy density of non-interacting electrons (Bader & Beddall, 1972), which can be represented as an explicit functional of the density in the framework of the orbital-free DFT. Specifically here we adopt the simple approximation based on the second order gradient expansion (Kirzhnits, 1957):

$$t_s(\mathbf{r}) \approx \frac{3}{10} (3\pi)^{2/3} \rho^{5/3}(\mathbf{r}) + \frac{1}{72} \frac{|\nabla \rho(\mathbf{r})|^2}{\rho(\mathbf{r})} + \frac{1}{6} \nabla^2 \rho(\mathbf{r}) \qquad (10)$$

The other semi-local approximations (Astakhov, Stash &. Tsirelson 2015) can be also used.

Equations (8)-(10) relate the electron pressure to the electron density, gradients of electron density and the Laplacian of electron density, which can be derived from accurate X-ray diffraction experiments. The pressure has dimension of Joule/m**3 and describes the self-organization of the quantum part of electronic energy density of a system in the position space (not electron density itself). Remarkably, even such a simple approximation for the distribution of quantum pressure allows locating the main chemical bonding features. It perfectly shows the electron shell structure, the covalent bonds and lone electron pairs as "compressed" (hard to the touch) regions of positive pressure, and the negative pressure "stretched" ("soft") regions of interatomic and intermolecular interactions as well as the ring and cage structural elements of molecules and crystals.

To conveniently map the distribution of the quantum electron pressure we used the scaled quantum electronic pressure-based function, Indicator of Quantum Pressure Focusing (IQPF) (Tao, Vignale & Tokatly, 2008)

$$L(\mathbf{r}) = \frac{1}{2}[1 + \frac{\gamma(\mathbf{r})}{\sqrt{1 + \gamma^2(\mathbf{r})}}]. \qquad (11)$$

Here $\gamma(\mathbf{r}) = p(\mathbf{r}) / p^{TF}(\mathbf{r})$, where $p(\mathbf{r}) = p^S(\mathbf{r}) + p^{xc}(\mathbf{r})$ is the actual quantum electron pressure normalized to the Thomas-Fermi pressure, that is, the pressure of the homogeneous electron gas evaluated at the actual local electron density: $p^{TF}(\mathbf{r}) = \frac{1}{5}(3\pi^2)^{2/3}\rho^{5/3}(\mathbf{r}) + p^{xc,LDA}(\mathbf{r})$. It is important to emphasize that IQPF by construction accounts both for the kinetic and the exchange-correlation contributions to the local electronic structure.

Similarly to the standard ELF[1], our IQPF is bounded to the interval $0 < L(\mathbf{r}) < 1$ with $L(\mathbf{r}) \to 1$ at $\gamma(\mathbf{r}) \to \infty$ and $L(\mathbf{r}) \to 0$ at $\gamma(\mathbf{r}) \to -\infty$. The value of $L = 1/2$ separates the compressed regions with a positive pressure $p(\mathbf{r}) > 0$, and the stretched regions with $p(\mathbf{r}) < 0$.

---

[1] The standard ELF is commonly used to extract the information about the spatial electron localization hidden in the electronic kinetic energy density. It is dimensionless scalar function which for closed-shell systems (Becke & Edgecombe, 1990; Savin, Becke, Flad, et al 1991; Silvi, & Savin, 1994; Grin, Savin & Silvi, 2014) takes forms

$$\eta(\mathbf{r}) = [1 + (D_P(\mathbf{r})/D_0(\mathbf{r}))^2]^{-1}.$$

$D_P(\mathbf{r}) = t_s(\mathbf{r}) - (1/8) |\nabla\rho(\mathbf{r})|^2/\rho(\mathbf{r})$ is Pauli kinetic energy density (March, 2010), $\rho(\mathbf{r})$ is the electron density and

$$t_s(\mathbf{r}) = (1/2)\sum_{l\sigma}|\nabla\psi_{l\sigma}(\mathbf{r})|^2$$

is single-determinant positively-defined electronic kinetic energy density; the summation is performed over all the occupied one-electron orbitals $\psi_{l\sigma}$. Function $D_0(\mathbf{r}) = (3/10)(3\pi^2)^{2/3}[\rho(\mathbf{r})]^{5/3}$ is the kinetic energy density of a uniform electron gas with density locally equal to $\rho(\mathbf{r})$. By construction, ELF is bounded to the interval $0 \le \eta \le 1$: regions with $\eta=1$ correspond to total electron pair localization, $\eta=1/2$ matches a uniform electron gas regions, and $\eta=0$ indicates the border between electron pairs. The zero-flux surfaces in the ELF gradient field enclose 3D basins which are associated with electron shells, bonds, or lone pairs. Each of basins contains an ELF attractor; basins are finite, disjoint, and space filling. Their sum gives the molecule or crystal cell volume. Some of pairs of attractors show a saddle point; low ELF value at this point is considered as a signature of the low electron delocalization between those basins.

In this work we used the experimental electron density and its derivatives expressed in terms of Hansen-Coppens (1978) multipole model.The actual values of all functions at each point **r** (the local homogeneity principle, see Kirzhnits, 1957) have been employed in computations. The core and valence radial functions of the multipole model were described by the atomic wavefunctions from (Macchi & Coppens, 2001). The multipole parameters for Si, $C_6H_6$, $NH_3$, $CaF_2$ and $\gamma$-$B_{28}$, crystalline compounds which present the typical chemical bonds, are taken from (Zuo, Blaha, & Schwarz, 1997; Burgi; Capelli, Goeta et al 2002; Boese, Niederpruem, Blaeser, et al 1997; Stachowicz, Malinska, Parafiniuk & Wozniak, 2017; Mondal, van Smaalen, Parakhonskiy, et al 2013), ), respectively. All the calculations have been performed using the program WinXPRO (Stash & Tsirelson, 2002, 2005, 2014), version 3.3.15.

### Results and Discussion

We have already demonstrated that combined application of the experimental electron density and the orbital-free DFT formalism opens the new avenue in quantum crystallography (Tsirelson, & Nagy, 2009; Tsirelson, Stash, & Liu, 2010; Tsirelson, Stash, Karasiev, Liu. 2013; Astakhov & Tsirelson, 2014; Astakhov, Stash, & Tsirelson 2016, Tsirelson, Stash & Tokatly, 2016). Nevertheless, we should explicitly state the restrictions of our approach resulting from using experimental electron density. The X-ray and synchrotron diffraction methods provide the diffracted intensities with precision of about 1% (Tsirelson & Ozerov, 1996; Herbst-Irmer & Stalke, 2017; Zhurov, Zhurova, & Pinkerton, 2008). The measurements are corrected for multiple scattering, thermal diffuse scattering, absorption and extinction, and the electron density of a crystal is reconstructed by means of the fit of the analytical multipole model to the obtained Bragg structure factors. Due to the model deficiencies (Koritsanszky, Volkov Chodkiewicz, 2012), limited resolution and incomplete thermal deconvolution (Madsen, 2012; Henn & Meindl, 2015), reconstructed electron density is distorted by the random and model errors, which are maximal in the vicinity of the nuclei. Therefore, these regions should be excluded from consideration. De-spite that, model electron density reasonably well reflects the Hartree-Fock and Kohn-Sham densities yielding the correct bonding picture in crystals with different types of the chemical bonds (Macchi, 2013). In addition, the second-order gradient expansion (10), in the framework of orbital-free DFT, shows acceptable local behavior of the electron kinetic energy density as well as approximate $p(\mathbf{r})$ (8) and IQPF (11) in the interatomic regions.

IQPF contains all information about the local quantum pressure and presents it in a compact, convenient for visualization form. Note that $L(\mathbf{r}) \geq 0.85$ if $p(\mathbf{r}) \geq p^{TF}(\mathbf{r})$ or $\gamma(\mathbf{r}) \geq 1$ and $L(\mathbf{r}) = 0.50$, if $p(\mathbf{r}) = 0$ ($\gamma(\mathbf{r}) = 0$). The Thomas-Fermi reference pressure built in the definition of IQPF makes evident the quantum origin of the electronic pressure focusing in the position space as compared to the homogeneous electron gas.

The IQPF maps computed using the approach outlined above are presented in the Fig. 1 - 5. Silicon (bcc, space gr. Fd3m) exemplifies s a crystal with typical non-polar covalent chemical bonds. The IQPF (Fig. 1) shows the areas of both maximal compression and stretching; we see the alternating closed regions of compression and stretching around Si nuclei (inner electron shells) slightly deformed in accordance

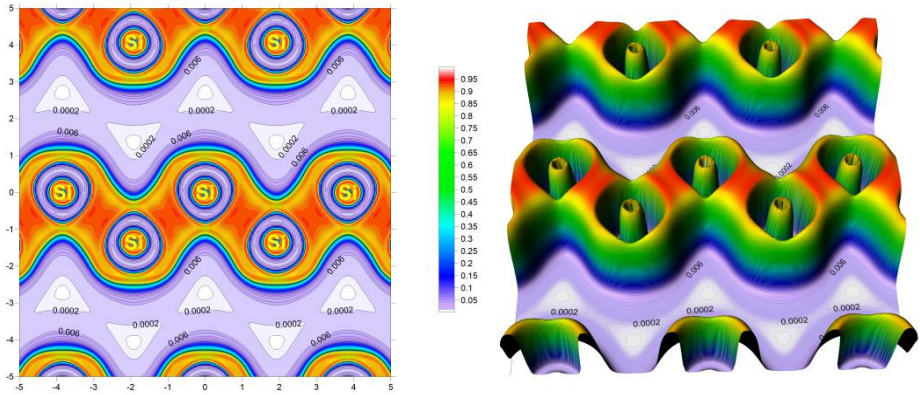

*Figure 1. Quantum pressure focusing indicator in the (110) plane of silicon. The characteristic values of 0.5 and 0.85 are indicated in the green and yellow colors*

with Td point symmetry of atomic positions. Along the interatomic Si−Si lines, the quantum pressure shows the large barrel-like space regions of strong electron compression. They are well-localized around the bond critical points in electron density (Abramov & Okamura, 1997). Due to quantum effects the pressure here is significantly higher than that of the uniform electron gas with the same density. Note that deformation electron density maps show the electron accumulation on the interatomic Si−Si lines (Tsirelson & Ozerov, 1996). The negative Laplacian of electron density (Abramov & Okamura, 1997) also signals about the electron density concentration in these regions. Small IQPF decrease is observed behind Si atoms on elongation of the Si−Si line, where deformation electron-density map shows the deficit and the Laplacian demonstrates the relative electron density depletion.

Out of the Si-Si bond lines, the pressure quickly decreases towards the octahedral (0, 0, ½) and tetrahedral (1/8, 1/8, 5/8) voids, where electron continuum strongly tends to stretch (softens) because of quantum effects. The electron gas is maximally stretched by quantum effects in tetrahedral voids. Note that electron density here is minimal and the most depleted ($\nabla^2\rho(\mathbf{r})$>0).

The crystal structure may be characterized by the gradient field of proper scalar function, not only of the electron density, as in QTAIM (Bader, 1990). For example, the IQPF attractors determine basins encompassing nuclei (atomic core basin) or valence basin, when no nucleus except a coreless H atom are present inside. The valence basins are classified by their connectivity or the synaptic order (Grin, Savin & Silvi, 2014 ), i.e. the number of core basins with which they share a common boundary. Monosynaptic basins are located behind atomic cores and contain the electron lone pairs, while di- and polysynaptic valence basins are between the two or more atomic cores and are referred to the chemical bonds. In the ELF theory, the closed space regions enveloped by the constant-valued surface are named localization domains (Silvi & Savin, 1994); they include one or more attractors. In this language, the IQPF identifies the covalent Si-Si bonds in silicon as the disynaptic valence basins of electronic continuum compression along Si–Si line.

Another example of the covalent bond characterization using IQPF provides the benzene molecule "extracted" from a crystal (Fig. 2). IQPF reveals the extended polysynaptic valence 0.85-domain of compressed electron gas (the surface spans the values IQPF with $L \geq 0.85$) containing all the C–C and C–H disynaptic valence basins. The picture reminds the "aromatic" domain enveloped by the surface ELF=0.5 in $C_6H_6$ as found early (Fuster, Sevin & Silvi, 2000). The C–C valence compression basin is elongated in the direction perpendicular to the molecule plane (manifestation of the $\pi$-component of bond – see insert in Fig. 2) and the C–H basin demonstrates

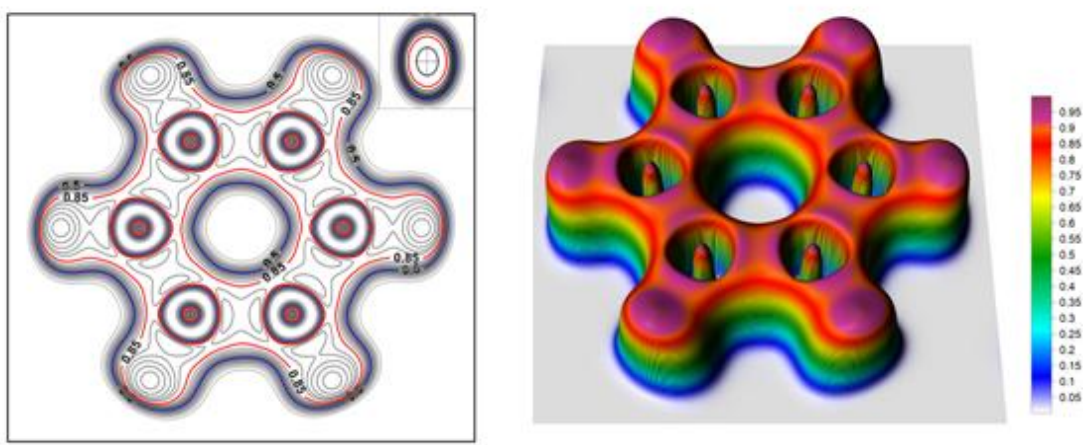

*Figure 2. 2D (left) and relief (right) maps of the quantum pressure focusing indicator for benzene molecule extracted from the crystal. The characteristic values of 0.5 and 0.85 are indicated by green and red colors. The insert shows the IQPF in the plane going perpendicular to the bond line at the middle of C-C distance*

the pressure shift towards more electronegative atom C. Note the visible compression of electrons within H atoms. The only region of the electron gas quantum stretch is located at the centre of an aromatic ring.

IQPF is able to reveal both bonding and lone electron pairs. It is seen on the ammonia molecule, Fig. 3, "extracted" from a crystal. The IQPF reconstructed from the experimental density reveals di- (bonds) and monosynaptic (electron lone pair) basins of the electron compression in $NH_3$ and shows certain electron delocalization between them.

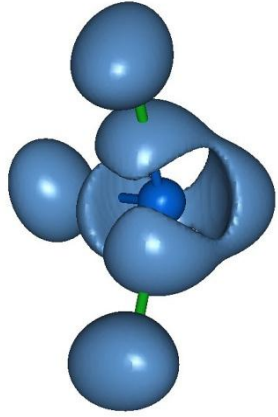

*Figure 3. 3D image of the quantum pressure focusing indicator for ammonia mole-cule extracted from the crystal. The IQPF domains of L=0.92 reveal the locations of bonding and lone-pair electrons.*

Our next example fluorite, $CaF_2$, crystallizes in a cubic face-centered lattice with a space group Fm-3m. Bonding in $CaF_2$ is typically described as ionic. Meanwhile, the bond path picture (it is not given here; also see Stachowicz et al, 2017) indicates more complex interaction pattern than it follows from classical crystal chemistry. Each Ca atom is linked to 8 F atoms; each F atom is tetrahedrally coordinated by atoms of Ca, and, additionally, is linked to 6 other F atoms. The F. …F interaction may be associated with so called charge-shift bond (Shaik, Danovich, Wu & Hiberty, 2009; Zhang, Ying, Wu, et al, 2009; Stachowicz et al, 2017) which is the resonance-energy related weak electron pair bonding with significant electrostatic component. Also, there is a crystal void centered at the point (0.5, 0.5, 0.5).

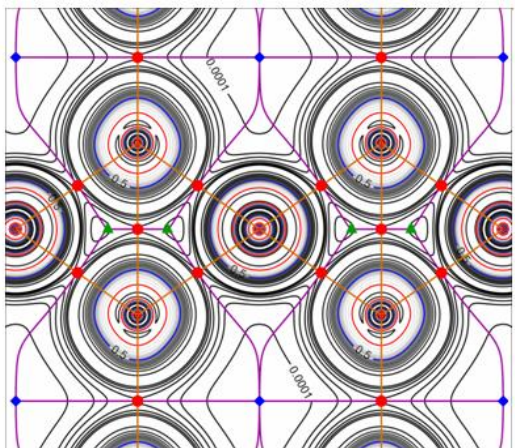

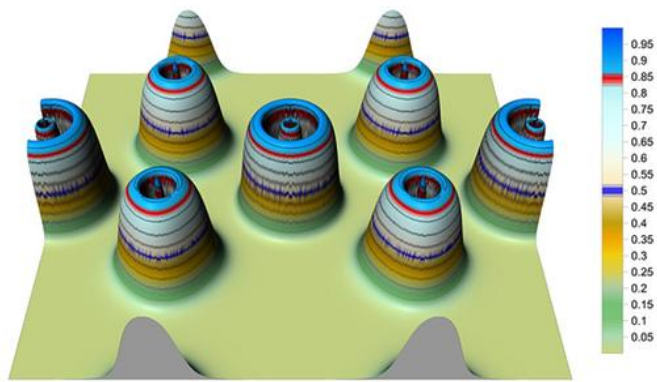

*Figure 4. Quantum pressure focusing indicator in the (110) plane of fluorite: 2D (a) and 3D(b) images. The2D map is superimposed with the bond path pattern from QTAIM. Red, green and blue circles denote the bond, ring and cage critical points in the electron density, correspondingly. The characteristic lines of 0.5 and 0.85 are indicated as well by the blue and red colors, respectively.*

IQPF, mapped in the (110) plane (Fig. 4), shows that the main variations of electronic continuum pressure in fluorite take place around the atomic positions. The inner electron shells of both atoms features alternating regions of quantum compression and stretch. The Ca atom is predominantly spherical, which is consistent with its ionic state with an net positive charge of +1.4 e. Fluorine ion occupies the volume comparable with that of Ca atom. Its position is described by non-centrosymmetric Td point symmetry and the small spatial asymmetry in the local pressure around the atoms F is clearly seen in Fig. 4.

The surface $L$=0.5 separates only atomic-like regions of quantum compression and stretching of electronic media. Disynaptic basins in $CaF_2$ are absent. The quantum electron pressure at the middle of ionic bond Ca–F and non-covalent interaction F...F is negative; it increases towards the nearby atoms of Ca and F, however, in ionic Ca–F bond, the anion is more extended along the bond line as compare with van der Waals F…F interaction. Probably, it reflects the electrostatic influence. Not unexpectedly, both these interactions correspond to areas of the positive Laplacian of electron density, i.e. to areas of electron density depletion (it is not shown here). It can be assumed that the periphery of atoms will be subject to largest deformation under external action. The most potentially deformable region of the fluorite crystal is located in the neighborhood of a crystal cell void.

Let us consider now the orthorhombic γ-phase of crystalline boron, $B_{28}$ (space gr. Pnnm). It is synthesized by compressing other boron phases to 12-20 GPa at 1500-1800°C (Zarechnaya, Dubrovinsky, Dubrovinskaia, et al, 2009; Oganov, Chen, Gatti, et al 2009) and remains stable at ambient temperature and pressure. The structure of γ-$B_{28}$ can be described as a rocksalt-type one consisting of the $B_{12}$ icosahedra (*closo*-cluster) and atomic pairs ("the dumbbell") connecting the icosahedral units. The bonds between pieces are described in the orbital language through two-center and three-center, two-electron bonding models (Macchi, 2011; Häussermann & Mikhaylushkin, 2010). Despite the fact that all the bonds of $\gamma$-$B_{28}$ are formed by the same kind of atoms, IQPF show for them different electron-density features.

First, the valence 0.5-domain, bounding the region of electron compression, has the polysynaptic character and is extended in the layer adjoined to the boron network in the crystalline unit cell. Only restricted media regions at the centre of $B_{12}$-clusters and between them are stretched. 0.85-domain shows also polysynaptic nature. Only at $L>0.9$ smaller additional localization domains appear in the boron network: they are mainly (but not necessary) located between the two atomic cores of boron atoms. We note that no bent "banana" B-B bonds have been found in $\gamma$-$B_{28}$.

Second, the bonds B(3)–B(3) linking the neighboring icosahedral clusters as well as B(5)–B(5) (the dumbbell) and B(2)–B(5) (the cluster-dumbbell) bonds are the most compressed in $\gamma$-$B_{28}$. Despite the difference in the length (1,660, 1,725, and 1,673 A respectively), they look very similar and do not show any evidence of the charge transfer from the cluster to dumbbell, as sometimes is assumed – see Macchi (2011) and Häussermann & Mikhaylushkin ( 2010) for discussion. The compression along B(2)–B(5) line is the weakest.

The maximally deformable region of the $\gamma$-$B_{28}$ crystal is located in the voids of the boron network, as expected.

## Conclusions

The proposed approach to recover the distribution of the electron pressure in the whole crystal space is compatible with a quantum mechanical description and significantly expands the framework of the accurate X-ray diffraction analysis. The high values of Indicator of Quantum Pressure Focusing (IQPF), $L(\mathbf{r})$, detects the regions of electronic continuum, which are locally compressed by the quantum effects in solids, both kinetic and interaction. By distinguishing the core electron shells and bonding and lone electron pairs, the electron pressure landscape allows also for categorizing the chemical bonds according to their types as it accepted in chemistry. Our analysis suggests that the regions with low internal electron pressure are the natural candidates to be more affected by the external influence. Our group is currently checking this assumption. Our approach also provides a natural tool to analyze the experimental electron density in crystals under pressure (Boldyreva, 2008; Casati, Kleppe, Jephcoat & Macchi, 2016; Casati, Genoni, Meyer et al, 2017; Katrusiak, 2008; Zakharov, Gal, Cruickshank & Boldyreva, 2018).

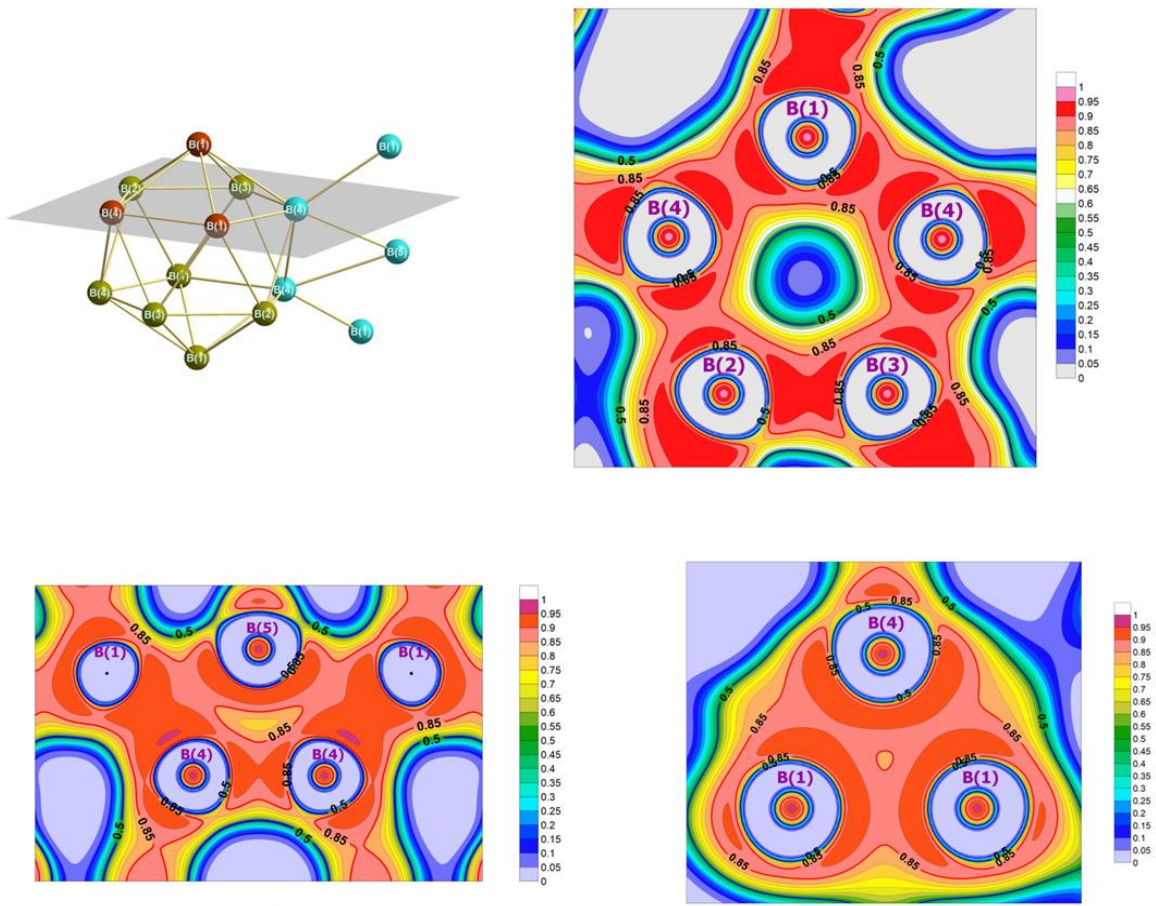

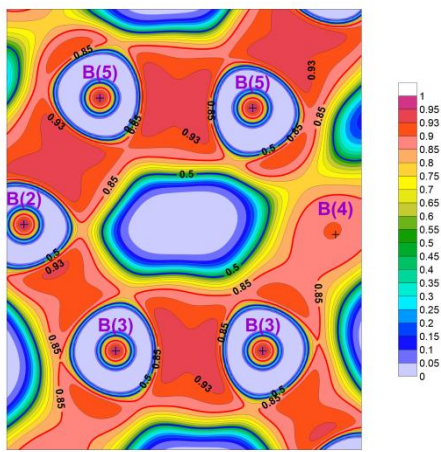

*Figure 5. Quantum pressure focusing indicator in the characteristic planes of γ-B28. The map in the plane through the B(1), B(4), B(2), B(3) and B(4) atoms is shown; the other planes are highlighted in the red and turquoise colors for atoms. The lines of L=0.5 (blue) and 0.85 (red) are indicated*

## Acknowlegements

The work of VGT and AIS is supported by the Russian Foundation for Basic Research, grant 16-03-00057a, the Government of the Russian Federation (Act 211, contract № 02.A03.21.0011) and by South Ural State University via grant of the Ministry of Education and Science of the Russian Federation, No 4.1157.2017/4.6. The work of IVT is supported by the Spanish Ministerio de Economía y Competividad (MINECO) Project No. FIS2016-79464-P and by the "Grupos Consolidados UPV/EHU del Gobierno Vasco" (Grant No. IT578-13).